\documentclass[10pt,a4paper]{article}
\usepackage{graphicx}
\usepackage{amsmath}
\usepackage{amssymb}
\usepackage[table]{xcolor}
\usepackage{booktabs}
\usepackage[T1]{fontenc}
\usepackage{lmodern}
\usepackage[margin=1in]{geometry}
\usepackage[utf8]{inputenc}
\usepackage{cite}
\usepackage{hyperref}
\usepackage{ragged2e}
\hypersetup{colorlinks,allcolors=black}
\usepackage[right]{lineno}
\usepackage{microtype}
\DisableLigatures[f]{encoding = *, family = * }
\justifying
\usepackage{changepage}
\usepackage[aboveskip=1pt,labelfont=bf,labelsep=period,singlelinecheck=off]{caption}
\makeatletter
\renewcommand{\@biblabel}[1]{\quad#1.}
\makeatother
\usepackage{lastpage,fancyhdr,graphicx}
\usepackage{epstopdf}
\usepackage{color}
\usepackage{multicol}

\begin{document}
\vspace*{0.35in}

\begin{flushleft}
{\Huge\textbf\newline{\textbf{Towards efficient structure prediction and pre-compensation in multi-photon lithography}}}\newline

Nicolas Lang\textsuperscript{1},
Sven Enns\textsuperscript{1},
Julian Hering\textsuperscript{1,2,*},
Georg von Freymann\textsuperscript{1,2,3}
\\
\bigskip
\textsuperscript{1} Technische Universit\"at Kaiserslautern TUK, Erwin-Schr\"odinger-Straße\,56, 67663 Kaiserslautern, Germany
\\
\textsuperscript{2} Opti-Cal GmbH, Erwin-Schr\"odinger-Straße\,56, 67663 Kaiserslautern, Germany
\\
\textsuperscript{3} Fraunhofer Institute for Industrial Mathematics ITWM, Fraunhofer-Platz\,1, 67663 Kaiserslautern, Germany
\\
\bigskip
*hering@physik.uni-kl.de

\end{flushleft}

\section{Abstract}
Microscale 3D printing technologies have been of increasing interest in industry and research for several years. Unfortunately, the fabricated structures always deviate from the respective expectations, often caused by the physico-chemical properties during and after the printing process. Here, we show first steps towards a simple, fast and easy to implement algorithm to predict the final structure topography for multi-photon lithography -- also known as Direct Laser Writing (DLW). The three main steps of DLW, (i) exposure of a photo resin, (ii) cross-linking of the resin, and (iii) subsequent shrinkage are approximated by mathematical operations, showing promising results in coincidence with experimental observations. E.g., the root-mean-square error ($rmse$) between the unmodified 3D print of a radial-symmetrically chirped topography and our predicted topography is only 0.46\,\textmu m, whereas the $rmse$ between this 3D print and its target is 1.49\,\textmu m. Thus, our robust predictions can be used prior to the printing process to minimize undesired deviations between the target structure and the final 3D printed structure. Using a Downhill-Simplex algorithm for identifying the optimal prediction parameters, we were able to reduce the $rmse$ from 4.04\,\textmu m to 0.33\,\textmu m by only two correction loops in our best-case scenario ($rmse=0.72$\,\textmu m after one loop). Consequently, this approach can eliminate the need for many structural optimization loops to produce highly conformal and high quality micro structures in the future.


\section{Introduction}
\label{sec:intro}
As one of the most flexible and high-resolution 3D printing technologies, multi-photon lithography a.k.a. Direct Laser Writing (DLW) has become established within the past 25 years. In a nutshell, a femto-second pulsed laser beam operating at near-infrared wavelength is tightly focused into a photo resin. There, a multi-photon absorption initiated polymerization takes place along a pre-programmed relative movement between focus and resin \cite{hohmann2015three,fischer2013three}. The technology's basics were laid by Maruo \textit{et al.} in 1997 \cite{maruo1997three}, whereas nowadays modern applications can be found in a wide range of disciplines: in integrated photonics, for example, the selective Bragg reflection band of liquid crystalline photo resins can be adjusted continuously in the visible range when applying DLW, paving the way for "true-color 3D" (or 4D) micro fabrication \cite{ritacco2022tuning}. Further applications can be found in life science for the fabrication and mimicking of 3D cellular micro-environments with tunable properties \cite{babi2021tuning}, in micro-optics for the fabrication of, e.g., Fresnel lenses onto fibers \cite{kumar2022emission}, in micro-mechanics \cite{stassi2021reaching}, micro-fluidics \cite{kunze2021taking} or in the context of topological photonics \cite{schulz2021topological,jorg2021observation}, to name just a few. Even in industrial research, DLW becomes more and more important, e.g., for providing ISO-conform calibration measures \cite{eifler2018calibration,dai2021define} or for industrial prototyping and mastering \cite{aderneuer2021two}. Especially, recent progresses towards direct laser writing of metals \cite{waller2021photosensitive} open up completely new possibilities in future.

Unfortunately, directly laser-written structures always exhibit some deformation compared to their original designs. These deformations can be observed either as shrinkage or bulging of the structures and occur both during and after the printing process. Since the density $\rho$ of a negative-tone photo resin's unpolymerized state is smaller than the density of the polymerized state \cite{zhou2015review}, there must be a loss in volume \textit{V} when the mass \textit{m} is conserved, following $\rho=m/V$. Here, shrinkage depends on the degree of polymerization and thus on the intensity of the laser used for printing. During the development of the printed structures, further shrinkage takes place due to the dissolution of soluble components. The latter include unpolymerized monomers and unreacted photoinitiators of the resin \cite{denning2011control}. Since the developer induces structure swelling in a slight amount, capillary effects during the drying process also lead to shrinkage \cite{meisel2006shrinkage}, which has been found to be the most prominent contributor \cite{sun2012situ}.

The aforementioned bulging of the printed structures is mainly caused by the so-called proximity effect \cite{waller2016spatio,saha2017effect,yang2019schwarzschild}. Here, the overlap of single voxels along the laser focus' trajectory while printing a structure overlap in space and time, leading to a local heightening of the exposure dose. According to this, e.g., a designed flat structure, like a disc, usually shows a curvature in its topography with a height maximum in its lateral center.

Both structure deforming aspects, the bulging and the shrinkage, cannot be avoided and ultimately result in structure deviations of up to 30\% \cite{purtov2018improved}. Fortunately, there are different ways for minimizing those deviations: (i) by pre-compensating the target shape prior to the printing, \cite{zhou2015review,sun2004shape,eifler2018calibration,meisel2006shrinkage,thiele20173d}, (ii) by spatially adjusting the laser dose during printing \cite{dai2021define,zhou2015review,aderneuer2021two}, or (iii) by improving the development process subsequent to the printing \cite{oakdale2016post,purtov2018improved}. While the latter does not take the proximity effect into account and is therefore fundamentally limited, the first two approaches aim for an overall homogeneous cross-linking of the resin and can theoretically achieve arbitrarily high structural conformations. In our case, spatially adjusting the laser dose requires significantly more computer storage, because a laser intensity value has to be added to each single coordinate within the programmed structure to be printed. This is especially important for large structures and can sum up to several additional GB, depending on the discretization. Although this approach allows for the smallest overall printing time, the calibration of how to adjust the laser's intensity in dependence of the respective structures is very challenging and has been patented recently \cite{tanguy2021method}.

Therefore, we rather focus on the pre-compensation of the target structure, which is usually designed on the computer, translated into coordinates for the DLW system and then printed. The printed structure is then measured, e.g. with a confocal microscope, and the thus accessible difference between the 3D print and its target is incorporated into a second version of the target structure. After printing this second version, its topography is measured again so that the target structure can be adjusted a second time. This process is repeated until the printed structure meets the respective requirements \cite{eifler2018calibration,thiele20173d}. Even though this approach leads to very high structural conformance, it is still time-consuming and labor-intensive, as five to ten (or even more) correction iterations are not uncommon \cite{eifler2018calibration}.

Predictions of the outcome of direct laser written structures reach back to first considerations on the expected widths of single lines by Fischer \textit{et. al} in 2013 \cite{fischer2013three-} or voxel and pillar dimensions by Purtov \textit{et. al} in 2019 \cite{purtov2019nanopillar}. Taking the reaction diffusion during the polymerization into account, dramatically complicates the modeling of structure prediction - even for simple lines, as it was recently published by Pingali and Saha \cite{pingali2022reaction}. There, it was only possible to reliably predict line widths, but not line heights or aspect ratios. Guney and Fedder published a promising semi-empirical analytic model through simulations and fitting for estimating widths and heights of single lines in 2016 \cite{guney2016estimation}. Unfortunately, this approach also cannot be easily transferred to voluminous structures of several tens or hundreds of microns length. Palmer \textit{et. al} recently focused on the simulation of additive manufactured metallic micro structures \cite{palmer2021simulation}. Although this fabrication method is based on direct laser writing, the photonic processes differ fundamentally in many points, hence, its simulation algorithm cannot be directly transferred to conventional direct laser writing of polymers. The most recent work in this field was published by Ad\~ao \textit{et. al} in 2022 \cite{adao2022two}. They established an algorithm to predict the resulting laser dose for the resin's exposure at each coordinate, taking experimental parameters like scan speed, laser power, and numerical aperture into account. The presented scanning electron microscope images of the fabricated wave-guide structures indicate an impressive similarity to their predictions, unfortunately, without providing a quantitative value. On closer inspection, however, one realizes that the shrinkage behaviour is not taken into account, hence, providing only a qualitative agreement, still showing deviations from the target structure. Those deviations are uncritical for the presented functionality of the fabricated waveguides, but can be crucial for, e.g., the functionality of diffractive elements or for determining the metrological characteristics of calibration measures. 

In contrast to this, we report on a fast computable and easy to implement algorithm that predicts the above mentioned undesired deformations for different types of structures. It takes shrinkage and proximity effects into account and is optimized for a quantitative high conformity of (bulky) target and printed topographies. This offers the possibility to directly pre-compensate the structures to be printed in order to achieve the highest possible conformity between the target and the 3D printed topography.

\section{Materials \& methods}
\label{sec:mm}
All micro-structures presented within this study were designed as 2D surface matrices. The pixel indices within the matrices correspond to lateral positions $(x,y)$, whereas the respective matrix entries represent the height values at those positions $z(x,y)$. Exporting these surface matrices to \textit{stl}-files allows for a common translation into coordinates for the DLW system, using the software \textit{Describe} (Nanoscribe GmbH \& Co. KG). If not explicitly described otherwise, we used \textit{Describe} to discretize the structures in equidistant axial planes with a so-called slicing distance of 0.1\,µm and each plane into lateral lines with a 0.1\,µm spacing -- called hatching distance. The thus generated data can be interpreted by the associated 3D printer \textit{Photonic Professional GT\textsuperscript{+}} (Nanoscribe GmbH \& Co. KG), using galvanometric mirrors for lateral, and a piezo stage for axial positioning. A constant laser power of roughly 40\,mW illuminating the complete entrance pupil of a 63x objective ($\text{NA}=1.4$ -- Carl Zeiss Microscopy Deutschland GmbH) and a constant writing speed of 20,000\,µm/s were used for all structures, as well as IP-S (Nanoscribe GmbH \& Co. KG) as photo resin. For all experiments we scanned the structures in an unidirectional way along the \textit{y}-axis and kept all the fabrication parameters (e.g., acceleration and deceleration of the galvo mirrors) constant. The development procedure after printing onto cleaned and silanized glass substrates \cite{liu20183d} followed the manufacturer's specifications: first, resting for 20 minutes in propylene glycol methyl ether acetate (PGMEA), second, resting for five minutes in isopropanol, and third, drying gently with nitrogen.

For measuring the structure's topographies, a \textit{µSurf} confocal microscope (NanoFocus AG) equipped with a 100x objective ($\text{NA}=0.95$) and a 60x objective ($\text{NA}=0.9$, Olympus Europa SE \& Co. KG, both) were used. To obtain the best possible results, the respective exposure times were optimized for each single measurement.

In complete analogy to the designed target structures, the data measured in this way correspond to 2D surface matrices. To be able to compare both data-sets, a correct alignment is of great importance: first, the non-measured points are interpolated using the bivariate spline function \textit{scipy.interpolate.BivariateSpline} of \textit{Python}'s \textit{SciPy} package \cite{2020SciPy}. Second, the target data is rescaled to fit the resolution of the measurement, utilizing the linear interpolation function \\
\textit{scipy.interpolate.RegularGridInterpolator} \cite{2020SciPy}. Thanks to the square imprints of all the structures, the measurement is then manually and eye-controlled rotated with respect to an illustrative vertical reference line that crosses the measurement field and to which the vertical edges of the structures are aligned in parallel. In this way, we align our measurement as perfectly as possible with the respective target surface, whereupon the data is cropped. Since both fabrication and measurement are not perfectly plane-parallel processes, a least-squares plane adjustment is used as a final pre-processing step to remove tilt.

The obtained data is used as input information for the algorithm presented below. Everything shown in the following was performed by a low-cost personal computer (PC) with only 4\,GB RAM, an Intel Pentium CPU G6950 processor with 2.80\,GHz and an Intel HD graphics card with 64\,GB VRAM. The algorithm's calculation times are less than one to five seconds excluding the optional parameter calibration routine (see section \ref{sec:pred}) and in the range of less than one to five minutes including the calibration, depending on the chosen discretization.

\section{The prediction algorithm}
\label{sec:pred}
\begin{figure}
\centering\includegraphics[width=\textwidth]{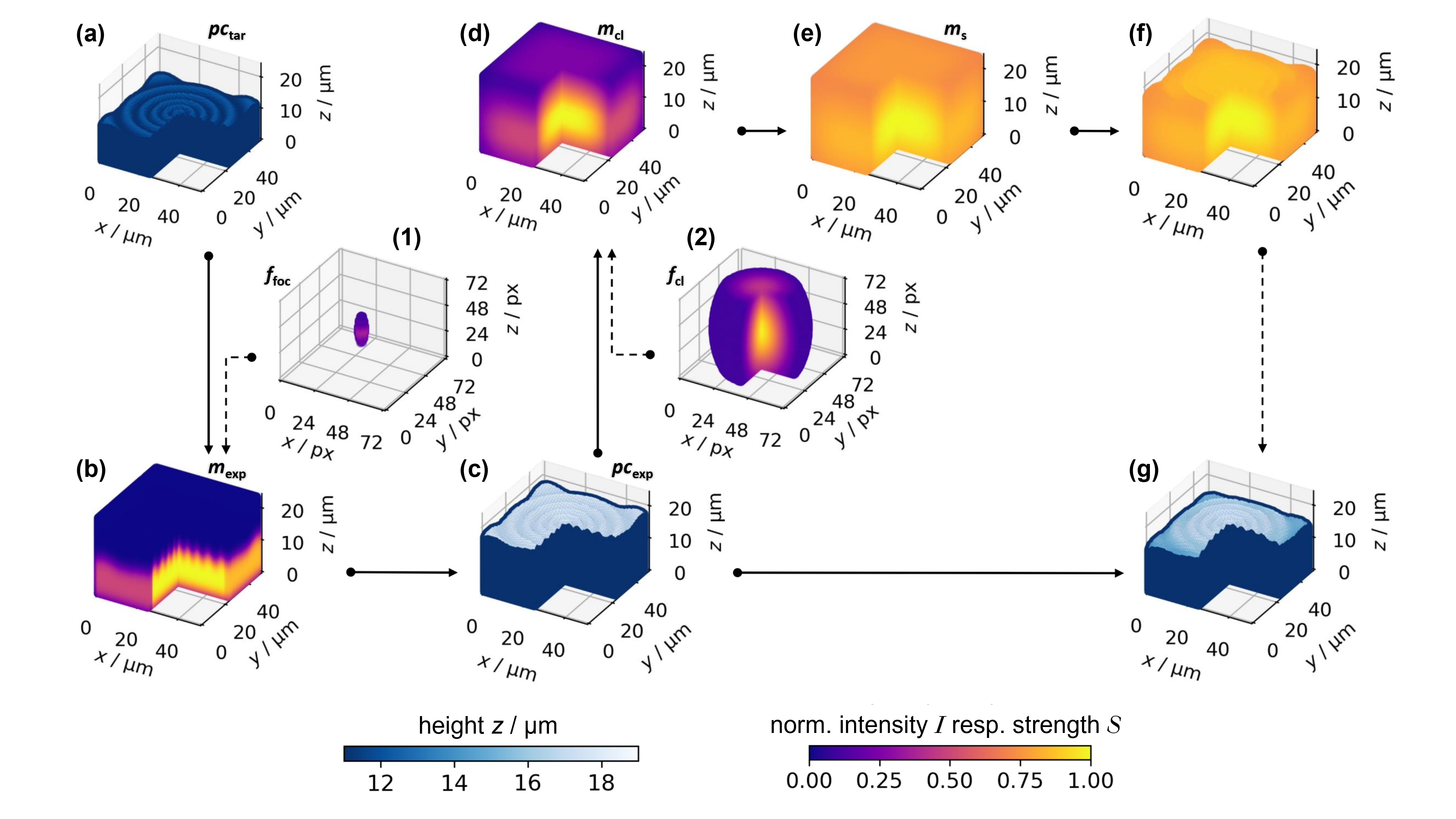}
\caption{Illustration of the presented prediction algorithm. All performed steps of the algorithm are schematically shown by the example of a radially chirped calibration measure. The final output (g) represents the finally expected 3D printed structure.}
\label{fig:scheme}
\end{figure}

In a first step the target structure is converted into a 3D point-cloud $pc_{\text{tar}}(x,y,z)$, containing the values 1 and 0, representing points within (1\,=\,True) and outside (0\,=\,False) the structure. This is schematically shown in Fig.\,\ref{fig:scheme}\,(a) as an overlay with a height-representing colormap. By adjusting the point-cloud's resolution, different hatching and slicing distances can be simulated as, e.g., doubling the lateral resolution of the point-cloud means halving the hatching distance.

Next, the spatial extent of the nearly Gaussian shaped laser focus is modeled as it moves through the photo resin during the printing process. Within this manuscript, we use a 3D Gaussian distribution $f_{\text{foc}}(x,y,z)$, illustrated as inset (1) in Fig.\,\ref{fig:scheme}, to represent the exposing laser focus. As we claim this to serve as 'first steps towards predictions', an independent adjustment of the lateral and axial expansions of the distribution was initially more important to us than the implementation of the theoretically correct point spread function. Moreover, the experimental laser focus always suffers from aberrations, typically leading to even stronger deviations from its theoretical shape, than those caused by our 3D Gaussian approximation \cite{hering2016automated}. The exposure of the resin is then approximated by convolving this Gaussian function with the structure point-cloud, resulting in the exposure matrix $m_{\text{exp}}(x,y,z)$:
\begin{equation}
    m_{\text{exp}} = pc_{\text{tar}} \circledast f_{\text{foc}},
    \label{eq:exposure}
\end{equation}
illustrated in Fig.\,\ref{fig:scheme}\,(b). Bisecting the exposure matrix in a part above and a part below a certain value allows the matrix to be binarized to values of 0 and 1, shown in Fig.\,\ref{fig:scheme}\,(c). This equals a threshold-dependent exposure point-cloud $pc_{\text{exp}}$ and assumes the existing minimum intensity for initializing cross-linking (compare threshold model \cite{yang2019schwarzschild,fischer2013three,fischer2013three-}).

As investigated by Waller \textit{et. al} \cite{waller2016spatio}, the diffusion of radicals and the cross-linking of monomers within the exposed resin can be approximated by a Gaussian distribution, whose full widths at half maximum (FWHM) defines the spatial extent. Hence, by calculating the convolution of $pc_{\text{exp}}$ with this second Gaussian distribution $f_{\text{cl}}(x,y,z)$ shown as inset (2) in Fig.\,\ref{fig:scheme}, the resulting cross-linking in 3D space within the resin is modeled
\begin{equation}
    m_{\text{cl}} = pc_{\text{exp}} \circledast f_{\text{cl}}
    \label{eq:cl}
\end{equation}
and shown in Fig.\,\ref{fig:scheme} (d). Subsequently, a linear mapping operation (equation \ref{eq:lm}) transfers this cross-linking matrix into a shrinkage matrix $m_{\text{s}}$, ascribing every point of $m_{\text{cl}}$ a specific shrinkage factor $\kappa$:
\begin{equation}
    m_{\text{s}} = \frac{\kappa_{\text{max}} - \kappa_{\text{min}}}{\text{max}(m_{\text{cl}}) - \text{min}(m_{\text{cl}})} \cdot m_{\text{cl}} + \kappa_{\text{min}},
    \label{eq:lm}
\end{equation}
with $\kappa_{\text{max}}$ and $\kappa_{\text{min}}$ being the maximal and minimal shrinkage factors, respectively. This linear mapping is illustrated in Fig.\,\ref{fig:scheme}\,(e) and includes the assumption of shrinkage factors being positive, non-zero and smaller than one to mathematically map the experimentally observed shrinkage. Typically, our algorithm is initiated with $\kappa_{\text{max}}=1$ and $\kappa_{\text{min}} = 0.7$. Due to the complexity of a holistic description of shrinkage in 3D, only the axial direction is considered directly here as an approximation, although the lateral influences due to cross-linking are of course also taken into account. This approach seems justifiable, since the previously mentioned deformations can be satisfactorily represented by local height adjustments for the rather simple 2.5D structures studied here, as it was already exploited for correction, e.g., in \cite{dai2021define}. As a price for this simplification we accept not only the limitation to 2.5D structures but also vertical edges within topographies being not deformed laterally, as is often observed experimentally. Accounting for true 3D shrinkage will be a future step, paving the way for the prediction of arbitrary 3D structures with subintersections or similar complexities.
Until this point, we average the values of $m_{\text{s}}$ along the \textit{z}-direction, leading to a 2D description:
\begin{equation}
    m_{\text{s}}^{\text{2D}}(x,y) = \frac{1}{N_{xy}}\sum_{z=0}^{N_{xy}}m_{\text{s}}(x,y,z),
    \label{eq:av}
\end{equation}
where $N_{xy}$ is the amount of $z$-values at the lateral position $(x,y)$ within the matrix. In the same fashion, $pc_{\text{exp}}$ is translated into a 2D matrix $pc_{\text{exp}}^{\text{2D}}$, representing the structure's height values as respective matrix entries. As shown in Fig.\,\ref{fig:scheme}\,(f), multiplying the obtained $z$-averaged shrinkage matrix $m_{\text{s}}^{\text{2D}}(x,y)$ with $pc_{\text{exp}}^{\text{2D}}$ only takes the shrinkage at the exposed positions into account and results in the final lithographic print prediction, which is illustrated as point-cloud in Fig.\,\ref{fig:scheme}\,(g).

Some of the parameters within this prediction algorithm are given by the experimental writing process itself and are therefore fixed, like hatching and slicing distances or the laser power. Other parameters, like the FWHM of the cross-linking distribution or the minimal and maximal shrinkage factors are not specifically accessible, since these parameters depend on the structures to be printed. Fortunately, all of these parameters represent real physical quantities whose range of values can be roughly identified by other experiments. For example, the FWHM of the laser focus can be estimated by fabricating and measuring individual voxels, or the range of influence of molecular diffusion can be estimated by the spatio-temporal controlled fabrication of single lines \cite{waller2016spatio}. Seven such physical-quantity based parameters are necessary for the prediction algorithm shown here:

\begin{itemize}
\setlength\itemsep{0em}
    \item the lateral and axial FWHM of the exposing Gaussian function\\
    (default values: $\sigma_{xy}^{\text{exp}}=0.5$\,\textmu m, $\sigma_{z}^{\text{exp}}=1.5$\,\textmu m),
    \item the polymerization threshold\\
    (default value: $P_{\text{thresh}}=50$\,\% of maximal power),
    \item the lateral and axial FWHM of the cross-linking distribution\\
    (default values: $\sigma_{xy}^{\text{cl}}=7.0$\,\textmu m, $\sigma_{z}^{\text{cl}}=7.0$\,\textmu m), and
    \item the minimal and maximal shrinkage factors\\
    (default values: $\kappa_{\text{min}}=0.7$, $\kappa_{\text{max}}=1.0$).
\end{itemize}

The default values are based on our experimental observations. Of course, these values are influenced by the structure to be printed and do not guarantee for accurate results. Therefore, we additionally use a Downhill-Simplex algorithm \cite{nelder1965simplex} (DSA) to determine the optimal simulation parameters.

\begin{figure}
\centering\includegraphics[width=\textwidth]{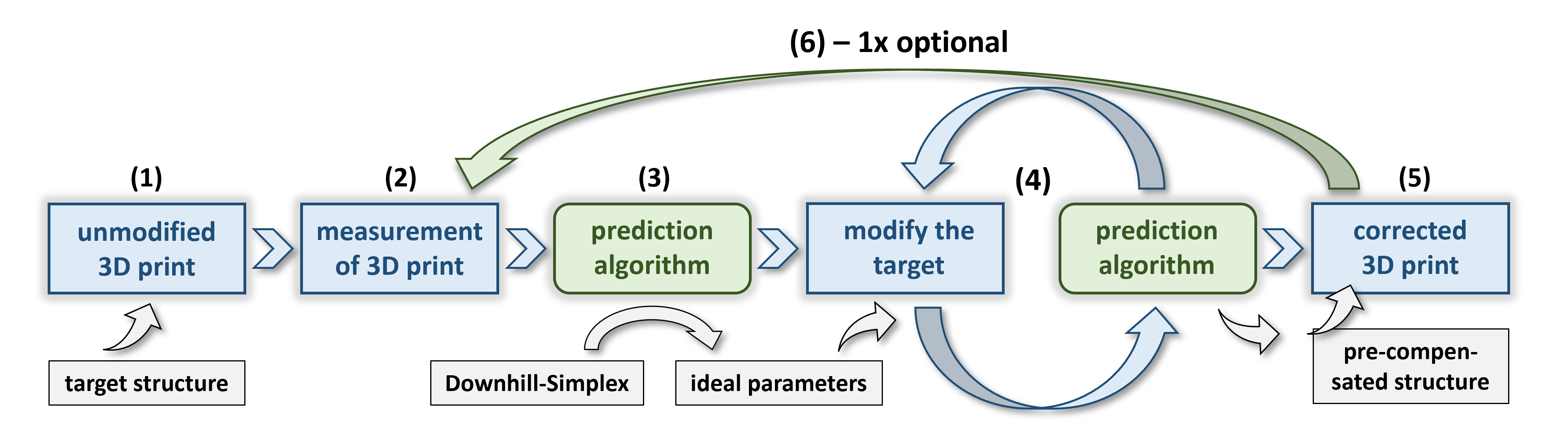}
\caption{Correction workflow. After printing (1) and measuring (2) a target structure, the presented prediction algorithm is used in combination with the Downhill-Simplex approach to identify the ideal prediction parameters (3). Subsequently, the latter are used for predicting again to iteratively modify the target structure until the prediction equals the very first target (4). Afterwards, the corresponding pre-compensated structure is printed (5). To further increase the conformity, one can optionally repeat steps (2) to (5) one time, represented by step (6).}
\label{fig:workflow}
\end{figure}

This algorithm generally minimizes the value of a given function with multiple variables. By determining the deviation between the prediction result and the corresponding measurement data of a structure as a function of the simulation parameters, the identification of the most appropriate simulation parameters is automated. Hence, one has to print and measure a first unmodified version of the target structure to identify those ideal simulation parameters by the DSA. On the one hand, this requires the printing of a so-called "calibration structure", on the other hand, the most suitable, structure-dependent simulation parameters are obtained. The related general workflow is illustrated in Fig.\,\ref{fig:workflow} and the quality of the respective results will be shown and discussed within the following section.

\section{Results \& discussion}
\label{sec:results}
Using the presented prediction algorithm for estimating the outcome of direct laser written structures requires the aforementioned set of simulation parameters. Without any measured data, one can only use educated guesses or use the default values given in section \ref{sec:pred}.

\begin{figure}
\centering\includegraphics[width=\textwidth]{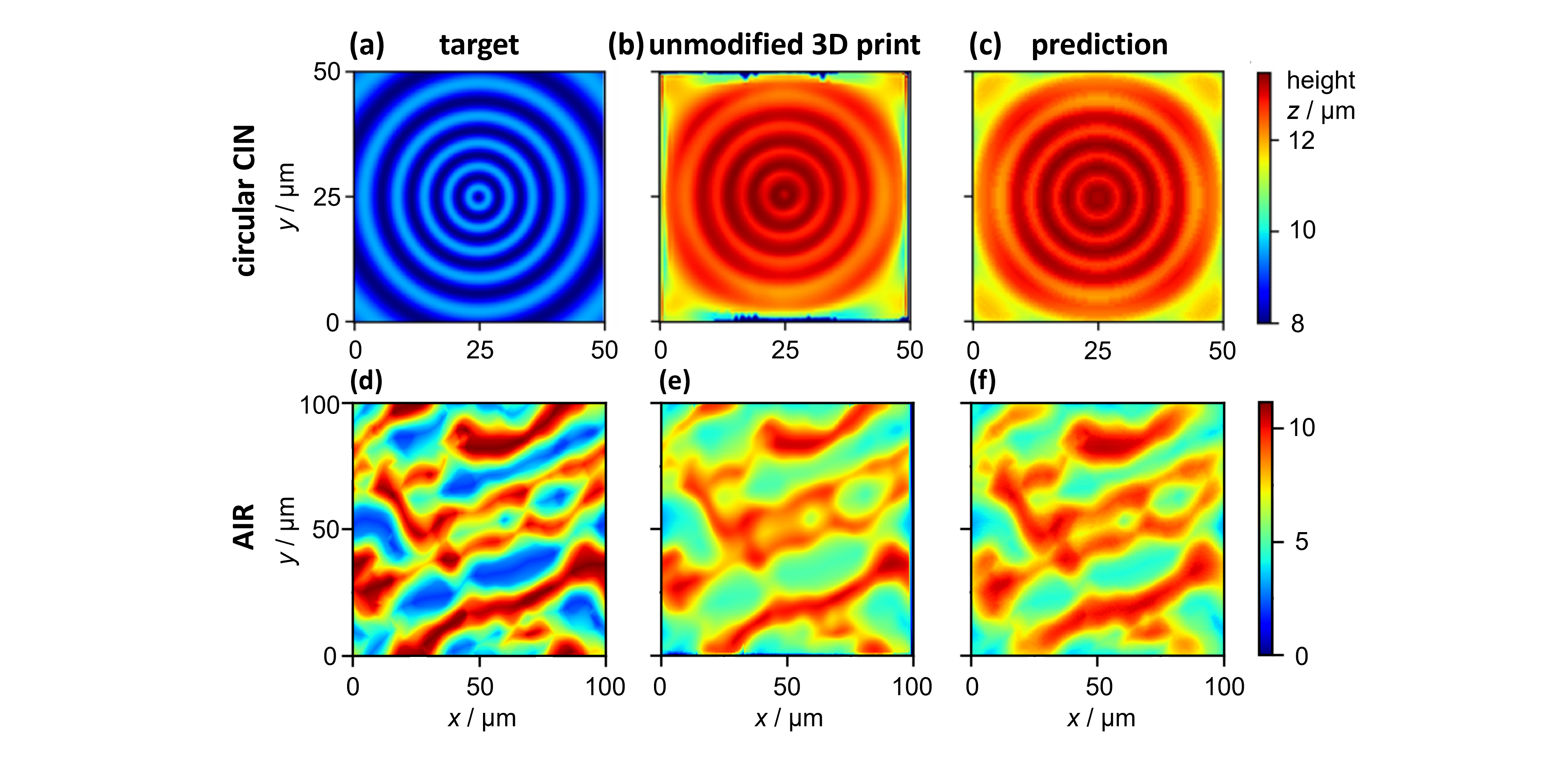}
\caption{Exemplary illustration of the prediction algorithm's performance. First, the described Downhill-Simplex approach based on the measurement of the unmodified 3D prints of a circular chirped (a \& b) and an areal roughness (d \& e) calibration measure is used to identify the structure-dependent ideal simulation parameters. Subsequently, those optimized parameters can be used for predicting the respective 3D prints (c) \& (f). This corresponds to the steps (1)\,-\,(3) of the workflow in Fig.\,\ref{fig:workflow}.}
\label{fig:pred}
\end{figure}

But if a calibration structure has been printed, it is recommended to use the parameters found by the Downhill-Simplex algorithm to predict the 3D printed surfaces (steps (1)\,-\,(3) of the workflow in Fig.\,\ref{fig:workflow}). The results of this approach are shown in Fig.\,\ref{fig:pred} for two exemplary types of structures: a circular chirped topography (CIN) and an areal roughness calibration measure (AIR). These two types of structures were specifically selected because they are widely used in the field of metrology: the AIR-type is based on an actual engineering surface and is designed with a tailored height distribution to provide a linear Abbot-curve. This is used for calibrating the height axis of measuring instruments as well as to determine their roughness properties \cite{eifler2018calibration}. The CIN-type, however, serves as resolution calibrating topography and is characterized by smooth topographic waves with an radially increasing wavelength. In contrast to other resolution-calibrating measures, the CIN is much less prone to measuring artifacts, due to the smooth waves. Additionally, it allows for determining the measuring instrument's transfer function in a continuous way and is not limited to specific axis orientations \cite{dai2021define}. Because of these metrological benefits and especially because of the accurate measurability due to the chosen amplitudes and topographic frequencies, height-deformations during and after the printing process are expected to be experimentally well observable. Hence, our predictions and pre-corrections should be well comparable, making AIR and CIN as test-structure very well suited.

In both cases, the predicted surfaces in Fig.\,\ref{fig:pred} (c) and (f) are pretty close to the actual measured ones in (b) and (e) and differ significantly -- as expected -- from their target shapes (a) and (d). To quantify the similarity between those topographies, one can use e.g., the root-mean-square error ($rmse$) and Matlab's 2D correlation coefficient \cite{Matlab}:
\begin{equation}
CC=\frac{\sum_m\sum_n(A_{mn}-\overline{A})(B_{mn}-\overline{B})}{\sqrt{\left(\sum_m\sum_n(A_{mn}-\overline{A})^2\right)\left(\sum_m\sum_n(B_{mn}-\overline{B})^2\right)}},  
\end{equation}
with $n$, $m$ being the indices of the matrices $A$, $B$, and $\overline{A}$, $\overline{B}$ representing the matrices' mean values. Hence, for the results shown in Fig.\,\ref{fig:pred} we get the first impression confirming values:
\begin{multicols}{4}
\begin{description}
\footnotesize{
\item $CC_{\text{(a),(b)}}^{\text{CIN}}=0.27$,
\item $CC_{\text{(a),(c)}}^{\text{CIN}}=0.39$,
\item $CC_{\text{(b),(c)}}^{\text{CIN}}=0.89$,
\item $rmse_{\text{(a),(b)}}^{\text{CIN}}=4.04\,\text{\textmu m}$,
\item $rmse_{\text{(a),(c)}}^{\text{CIN}}=3.73\,\text{\textmu m}$,
\item $rmse_{\text{(b),(c)}}^{\text{CIN}}=0.46\,\text{\textmu m}$,
\item $CC_{\text{(d),(e)}}^{\text{AIR}}=0.85$,
\item $CC_{\text{(d),(f)}}^{\text{AIR}}=0.88$,
\item $CC_{\text{(e),(f)}}^{\text{AIR}}=0.89$,
\item $rmse_{\text{(d),(e)}}^{\text{AIR}}=1.49\,\text{\textmu m}$,
\item $rmse_{\text{(d),(f)}}^{\text{AIR}}=1.60\,\text{\textmu m}$,
\item $rmse_{\text{(e),(f)}}^{\text{AIR}}=0.93\,\text{\textmu m}$.
}
\end{description}
\end{multicols}
The origin of the observed deviations between 3D print and target has already been explained in the introduction, but the prediction algorithm maps them quite well, as represented, e.g., by root mean square errors of 0.46 µm and 0.93 µm, respectively. Not only the total offset along the axial dimension due to the elongation of the exposing laser focus, but also the lift towards the center of the structure due to vignetting and the proximity effect are well captured. Note, that doubling the footprint of the structures from 50\,\textmu m of the CIN to 100\,\textmu m of the AIR does not worsen the algorithm's performance.

Due to those promising prediction capabilities, we use our approach for pre-corrections, illustrated by steps (4) and (5) of Fig.\,\ref{fig:workflow}. Ideally, the algorithm allows to modify the target structure so that the resulting 3D printed result matches the original target. Note that this process requires only two prints in total: the unmodified "calibration structures" and the final corrected structures.
\begin{figure}
\centering\includegraphics[width=\textwidth]{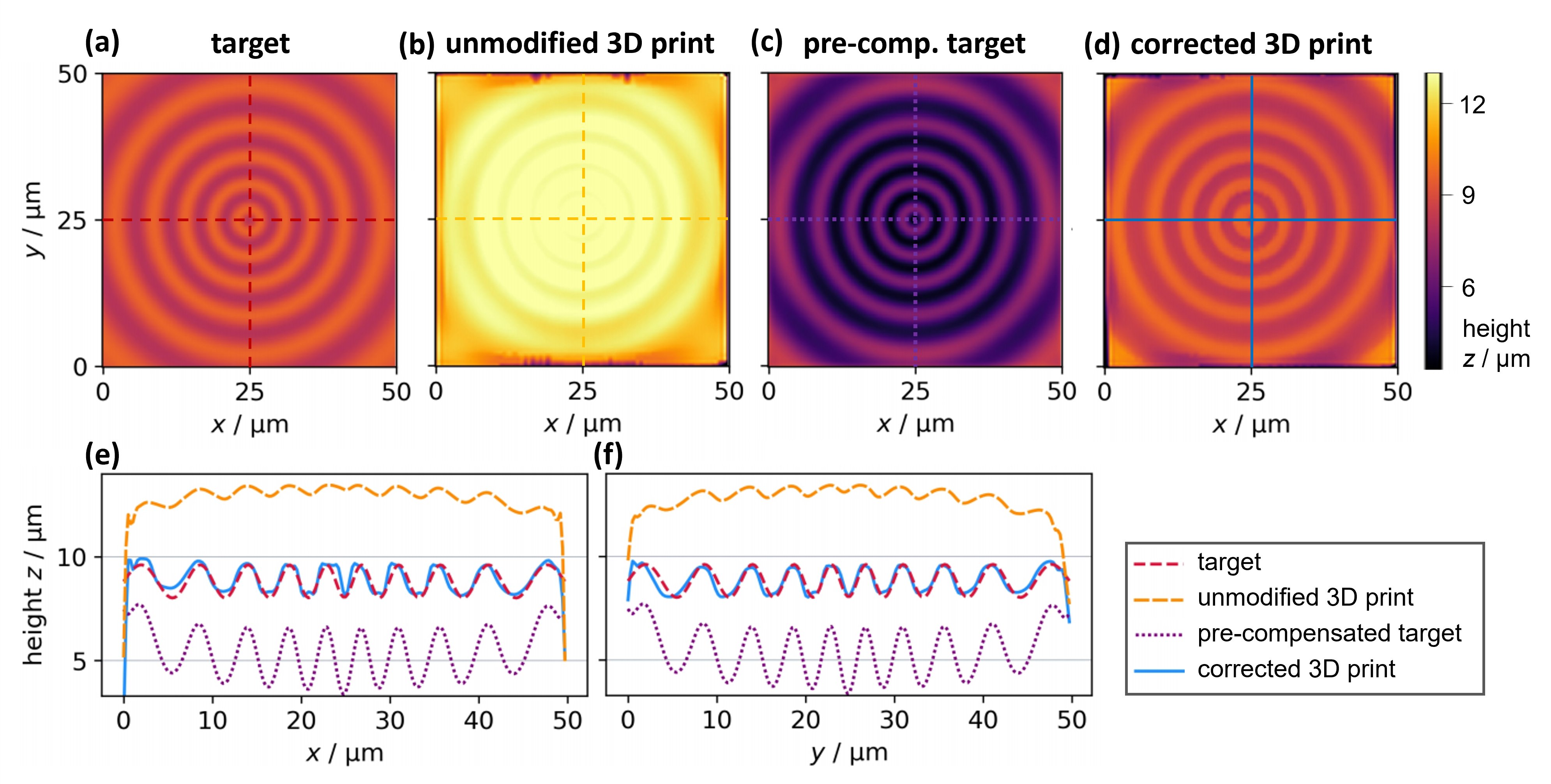}
\caption{Second generation correction of a circular chirped structure (CIN). The target topography (a) gets -- based on its measured surface (b) -- modified in such a way (c), that the final 3D print (d) is almost equal to the original target. The respective cross-sections are shown as $x$- (e) and $y$-profiles (f). This correction method represents the complete workflow, illustrated in Fig.\,\ref{fig:workflow}.}
\label{fig:corrCIN}
\end{figure}

However, the prediction of the modified structures within the loop of step (4) will be less accurate, since the simulation parameters have been optimized for the unmodified 3D print. For instance, the $rmse$ between the 'old parameter predicted' modified CIN structure and it's corresponding 3D print is 5.058\,\textmu m, confirming the aforementioned suspicion. To compensate for this, the described process can be repeated using the 'first generation' of correction in step (5) for a second run -- the optional but recommended step (6). This of course increases the number of total prints by one and is referred to as the 'second generation' of correction. We observed that the best results are achieved by those second generation structures (compare table\,\ref{tab:metrol}). Three printing processes are still much less compared to up to ten or even more, often required elsewhere \cite{eifler2018calibration}.

\begin{figure}
\centering\includegraphics[width=\textwidth]{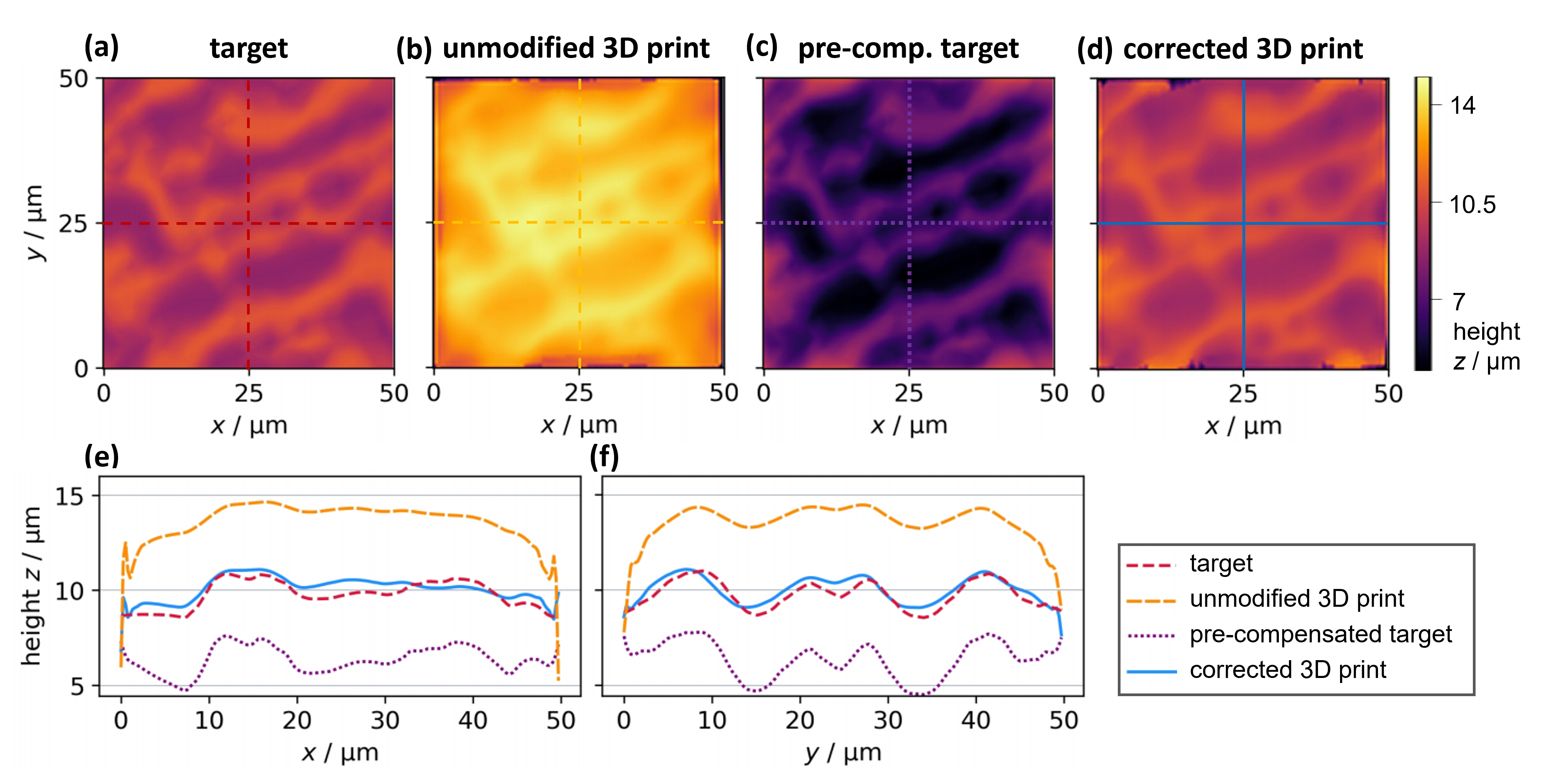}
\caption{Second generation correction of an areal roughness structure (AIR). The target topography (a) gets -- based on its measured surface (b) -- modified in such a way (c), that the final 3D print (d) is almost equal to the original target. The respective cross-sections are shown as $x$- (e) and $y$-profiles (f). This correction method represents the complete workflow, illustrated in Fig.\,\ref{fig:workflow}.}
\label{fig:corrAIR}
\end{figure}

Two exemplary second generation correction results for (i) a circular chirped (CIN) and (ii) an areal roughness (AIR) structure are shown in Fig.\,\ref{fig:corrCIN} and Fig.\,\ref{fig:corrAIR}, respectively.
The measured final 3D prints (d) are characterized by a very high conformity to their target surfaces (a), which can be seen nicely within the profile plots (e) \& (f). The deviations between the finally produced topographies and the respective targets can be greatly reduced. In terms of the root-mean-square error ($rmse$), these deviations decrease from 4.044\,\textmu m to 0.332\,\textmu m for CIN and from 3.490\,\textmu m to 0.477\,\textmu m for AIR, respectively. Besides, the values for the first generation corrections, as well as the corresponding 2D correlation coefficients are given in Tab.\,\ref{tab:metrol} and quantitatively confirm the improvements.

As these two types of structures are supposed to image specific metrological characteristics, we can also have a look at, e.g., the axial amplification coefficient ($\alpha_{z}$) or the quadratic areal roughness parameter $S_{\text{q}}$ of the AIR structure (see references \cite{ISO25178-603,eifler2018calibration,ISO25178-70} for details about the metrology).
\begin{table}[b]
\small
\caption{Improvement over correction generations. The 2D correlation coefficients $CC$ and root-mean-square errors $rmse$ for the structures depicted in Figs.\,\ref{fig:corrCIN} and \ref{fig:corrAIR} are shown for each correction generation. The values are calculated with reference to their respective target topographies. Moreover, the metrological characteristics of the AIR structure (axial amplification coefficient $\alpha_{z}$, linearity deviation $l_{z}$, arithmetic $S_{\text{a}}$, and quadratic $S_{\text{q}}$ areal roughness) are exemplary compared, too.} 
\centering
\begin{tabular}{lcccc}
\toprule
 & \textbf{target} & \textbf{calibration structure} & \textbf{1\textsuperscript{st} generation} & \textbf{2\textsuperscript{nd} generation}\\
\midrule
$CC^{\text{CIN}}$ & 1.000 & 0.270 & 0.686 & 0.848 \\
$CC^{\text{AIR}}$ & 1.000 & 0.476 & 0.762 & 0.830 \\
$rmse^{\text{CIN}}$ / \textmu m & 0.000 & 4.044 & 0.715 & 0.332 \\
$rmse^{\text{AIR}}$ / \textmu m & 0.000 & 3.490 & 1.270 & 0.477 \\
$\alpha_{z}$ / 1 & 1.000 & 0.772 & 0.793 & 0.924 \\
$l_{z}$ / \textmu m & 0.000 & 0.196 & 0.036 & 0.036 \\
$S_{\text{a}}$ / \textmu m & 0.615 & 0.570 & 0.506 & 0.581 \\
$S_{\text{q}}$ / \textmu m & 0.710 & 0.741 & 0.596 & 0.695 \\
\bottomrule
\end{tabular}
\label{tab:metrol}
\end{table}

As shown in Tab.\,\ref{tab:metrol}, the amplification coefficient for the unmodified calibration structure deviates by roughly 23\%, getting slightly improved to 21\% by the first generation correction. The second generation, on the other hand significantly improves this metrological characteristic towards only 8\% deviation. As a second example, the $S_{\text{q}}$ deviates by 30\,nm, by 110\,nm, and by 15\,nm through the different correction generations. Similar behaviours can be observed for the other metrological characteristics, as well as for the generally quantifying $rmse$ and $CC$ values (see. Tab.\,\ref{tab:metrol}), once again underlining the overall conformity enhancement, obtained by the presented method.

This can be additionally strengthened by the convergence of the simulation parameters by the Downhill-Simplex algorithm. There, the FWHM of the exposing laser focus changes from the default values ($\sigma_{xy}^{\text{exp}}=0.5$\,\textmu m, $\sigma_{z}^{\text{exp}}=1.5$\,\textmu m) to roughly 0.41\,\textmu m and 2.15\,\textmu m, respectively for the 2\textsuperscript{nd} generation AIR-type structure. Hence, the aspect ratio of the simulated voxel increases from 1.5 (default) to 5.2, which is based on the manufacturer's specifications much more realistic for this full-volume structure and the photo resin used \cite{NanoGuide}. To just name a second example, the spatial cross-linking slightly decreases from $\sigma_{xy}^{\text{cl}}=\sigma_{z}^{\text{cl}}=7$\,\textmu m (default) to values between 4.4\,\textmu m and 5.7\,\textmu m for both AIR and CIN, being more realistic, too, following the calculations of reference \cite{waller2016spatio}. An analogous behaviour can be observed for all 2\textsuperscript{nd} generation correction simulation parameters, demonstrating the power of our approach.

\section{Summary \& outlook}
In this study, we have presented an approach to predict the topography of directly laser-written structures. Our algorithm considers several physical quantities as simulation parameters to account for the main physico-chemical processes during fabrication that are responsible for undesired deviations. In addition to fixed parameters, such as hatching and slicing distances or laser power, parameters that are difficult to access, such as the effective exposure or the spatial region of cross-linking in the photo resin, can be automatically optimized for each structure. Since the resulting 3D printing predictions are very promising (e.g., $rmse$ reduced by more than 4\,\textmu m down to be less than 1\,\textmu m), an iterative application of the algorithm allows a reasonable pre-compensation of the structures to be printed. In the end, one can expect a very high match between target and printed structure within only two or optionally three printing steps. This match is for instance represented by the 2D correlation coefficient being 0.27 and 0.48 for the unmodified CIN and AIR structures, respectively. The first correction generation improves these values to 0.69 and 0.76, whereas the third generation even further enhances them to 0.85 and 0.83.

Since, e.g., the properties of the used photo resin or the size of the point spread function can be seen as covered by the automatically optimized prediction parameters, our approach should be adaptable to other resins, objectives, and different kinds of structures. In contrast to that flexibility, we are fundamentally limited to 2.5D structures right now due to the mathematical working principle of the prediction algorithm. However, for many applications 2.5D structures are sufficient, as micro-lenses, Fresnel-lenses, diffraction gratings and prisms belong all to this class.

An extensive investigation in terms of topographical, material as well as process capabilities will provide further insight into the limitations of our approach but exceeds the claim of 'first steps' aimed at with this publication. As further future work, we will speed up the identification of optimal simulation parameters. Conceivable here would be a neural network trained by our prediction algorithm that can set the optimal simulation parameters for each generation. Ideally, even without the need to print any calibration structures beforehand. For this purpose, we will extend our prediction algorithm to take into account the holistic shrinkage behavior in 3D to predict arbitrary complex 3D structures, as well as the deformation of vertical edges.
\\
\\
\noindent{\bfseries Acknowledgement.}
{\small Funded by the Deutsche Forschungsgemeinschaft (DFG, German Research Foundation) – Project-ID 172116086 – SFB 926. The authors gratefully acknowledge the institute for measurement and sensor technology (MTS) at the University of Kaiserslautern for the opportunity of taking confocal measurements.}
\\
\\
\noindent{\bfseries Disclosures.}
{\small The authors declare no conflicts of interest.}
\\
\\
\noindent{\bfseries Data availability.} 
{\small The data used to support the findings of this study are available from the corresponding author
upon request.}

\nolinenumbers

\bibliography{library}

\begin{thebibliography}{10}

\bibitem{hohmann2015three}
Judith~K Hohmann, Michael Renner, Erik~H Waller, and Georg von Freymann.
\newblock Three-dimensional $\mu$-printing: An enabling technology.
\newblock {\em Advanced Optical Materials}, 3(11):1488--1507, 2015.

\bibitem{fischer2013three}
Joachim Fischer and Martin Wegener.
\newblock Three-dimensional optical laser lithography beyond the diffraction
  limit.
\newblock {\em Laser \& Photonics Reviews}, 7(1):22--44, 2013.

\bibitem{maruo1997three}
Shoji Maruo, Osamu Nakamura, and Satoshi Kawata.
\newblock Three-dimensional microfabrication with two-photon-absorbed
  photopolymerization.
\newblock {\em Optics letters}, 22(2):132--134, 1997.

\bibitem{ritacco2022tuning}
Tiziana Ritacco, Dante~M Aceti, Gianfranco De~Domenico, Michele Giocondo,
  Alfredo Mazzulla, Gabriella Cipparrone, and Pasquale Pagliusi.
\newblock Tuning cholesteric selective reflection in situ upon two-photon
  polymerization enables structural multicolor 4d microfabrication.
\newblock {\em Advanced Optical Materials}, page 2101526, 2022.

\bibitem{babi2021tuning}
Mouhanad Babi, Roberto Riesco, Louisa Boyer, Ayodele Fatona, Angelo Accardo,
  Laurent Malaquin, and Jose Moran-Mirabal.
\newblock Tuning the nanotopography and chemical functionality of 3d printed
  scaffolds through cellulose nanocrystal coatings.
\newblock {\em ACS Applied Bio Materials}, 4(12):8443--8455, 2021.

\bibitem{kumar2022emission}
Aashutosh Kumar, Asa Asadollahbaik, Jeongmo Kim, Khalid Lahlil, Simon Thiele,
  Alois~M. Herkommer, S\'{i}le~Nic Chormaic, Jongwook Kim, Thierry Gacoin,
  Harald Giessen, and Jochen Fick.
\newblock Emission spectroscopy of nayf 4: Eu nanorods optically trapped by
  fresnel lens fibers.
\newblock {\em Photonics Research}, 10(2):332--339, 2022.

\bibitem{stassi2021reaching}
Stefano Stassi, Ido Cooperstein, Mauro Tortello, Candido~Fabrizio Pirri, Shlomo
  Magdassi, and Carlo Ricciardi.
\newblock Reaching silicon-based nems performances with 3d printed
  nanomechanical resonators.
\newblock {\em Nature Communications}, 12(1):1--9, 2021.

\bibitem{kunze2021taking}
Fynn~L Kunze, Torsten Henning, and Peter~J Klar.
\newblock Taking internally wetted capillary electrospray emitters to the
  sub-ten-micrometer scale with 3d microlithography.
\newblock {\em AIP Advances}, 11(10):105315, 2021.

\bibitem{schulz2021topological}
Julian Schulz, Sachin Vaidya, and Christina J{\"o}rg.
\newblock Topological photonics in 3d micro-printed systems.
\newblock {\em APL Photonics}, 6(8):080901, 2021.

\bibitem{jorg2021observation}
Christina J{\"o}rg, Sachin Vaidya, Jiho Noh, Alexander Cerjan, Shyam Augustine,
  Georg von Freymann, and Mikael~C. Rechtsman.
\newblock Observation of quadratic (charge-2) weyl point splitting in
  near-infrared photonic crystals.
\newblock {\em Laser \& Photonics Reviews}, 16(1):2100452, 2022.

\bibitem{eifler2018calibration}
Matthias Eifler, Julian Hering, Georg Von~Freymann, and J{\"o}rg Seewig.
\newblock Calibration sample for arbitrary metrological characteristics of
  optical topography measuring instruments.
\newblock {\em Optics Express}, 26(13):16609--16623, 2018.

\bibitem{dai2021define}
Gaoliang Dai, Xiukun Hu, Julian Hering, Matthias Eifler, J{\"o}rg Seewig, and
  Georg von Freymann.
\newblock Define and measure the dimensional accuracy of two-photon laser
  lithography based on its instrument transfer function.
\newblock {\em Journal of Physics: Photonics}, 3(3):034002, 2021.

\bibitem{aderneuer2021two}
Tamara Aderneuer, Oscar Fern{\'a}ndez, and Rolando Ferrini.
\newblock Two-photon grayscale lithography for free-form micro-optical arrays.
\newblock {\em Optics Express}, 29(24):39511--39520, 2021.

\bibitem{waller2021photosensitive}
Erik~Hagen Waller, Julian Karst, and Georg von Freymann.
\newblock Photosensitive material enabling direct fabrication of filigree 3d
  silver microstructures via laser-induced photoreduction.
\newblock {\em Light: Advanced Manufacturing}, 2(2):228--233, 2021.

\bibitem{zhou2015review}
Xiaoqin Zhou, Yihong Hou, and Jieqiong Lin.
\newblock A review on the processing accuracy of two-photon polymerization.
\newblock {\em Aip Advances}, 5(3):030701, 2015.

\bibitem{denning2011control}
Robert~G Denning, Christopher~F Blanford, Henning Urban, Harpal Bharaj, David~N
  Sharp, and Andrew~J Turberfield.
\newblock The control of shrinkage and thermal instability in su-8 photoresists
  for holographic lithography.
\newblock {\em Advanced Functional Materials}, 21(9):1593--1601, 2011.

\bibitem{meisel2006shrinkage}
Daniel~Christoph Meisel, Marcus Diem, Markus Deubel, Fabi{\'a}n
  P{\'e}rez-Willard, Stefan Linden, Dagmar Gerthsen, Kurt Busch, and Martin
  Wegener.
\newblock Shrinkage precompensation of holographic three-dimensional
  photonic-crystal templates.
\newblock {\em advanced materials}, 18(22):2964--2968, 2006.

\bibitem{sun2012situ}
Quan Sun, Kosei Ueno, and Hiroaki Misawa.
\newblock In situ investigation of the shrinkage of photopolymerized
  micro/nanostructures: the effect of the drying process.
\newblock {\em Optics letters}, 37(4):710--712, 2012.

\bibitem{waller2016spatio}
Erik~Hagen Waller and Georg Von~Freymann.
\newblock Spatio-temporal proximity characteristics in 3d $\mu$-printing via
  multi-photon absorption.
\newblock {\em Polymers}, 8(8):297, 2016.

\bibitem{saha2017effect}
Sourabh~K. Saha, Chuck Divin, Jefferson~A. Cuadra, and Robert~M. Panas.
\newblock {Effect of Proximity of Features on the Damage Threshold During
  Submicron Additive Manufacturing Via Two-Photon Polymerization}.
\newblock {\em Journal of Micro and Nano-Manufacturing}, 5(3):031002, 05 2017.

\bibitem{yang2019schwarzschild}
Liang Yang, Alexander M{\"u}nchinger, Muamer Kadic, Vincent Hahn, Frederik
  Mayer, Eva Blasco, Christopher Barner-Kowollik, and Martin Wegener.
\newblock On the schwarzschild effect in 3d two-photon laser lithography.
\newblock {\em Advanced Optical Materials}, 7(22):1901040, 2019.

\bibitem{purtov2018improved}
Julia Purtov, Andreas Verch, Peter Rogin, and Ren{\'e} Hensel.
\newblock Improved development procedure to enhance the stability of
  microstructures created by two-photon polymerization.
\newblock {\em Microelectronic Engineering}, 194:45--50, 2018.

\bibitem{sun2004shape}
Hong-Bo Sun, Tooru Suwa, Kenji Takada, Remo~Proietti Zaccaria, Moon-Soo Kim,
  Kwang-Sup Lee, and Satoshi Kawata.
\newblock Shape precompensation in two-photon laser nanowriting of photonic
  lattices.
\newblock {\em Applied physics letters}, 85(17):3708--3710, 2004.

\bibitem{thiele20173d}
Simon Thiele, Kathrin Arzenbacher, Timo Gissibl, Harald Giessen, and Alois~M
  Herkommer.
\newblock 3d-printed eagle eye: Compound microlens system for foveated imaging.
\newblock {\em Science advances}, 3(2):e1602655, 2017.

\bibitem{oakdale2016post}
James~S Oakdale, Jianchao Ye, William~L Smith, and Juergen Biener.
\newblock Post-print uv curing method for improving the mechanical properties
  of prototypes derived from two-photon lithography.
\newblock {\em Optics express}, 24(24):27077--27086, 2016.

\bibitem{tanguy2021method}
Yann Tanguy and Nicole Lindenmann.
\newblock Method for producing a 3d structure by means of laser lithography,
  and corresponding computer program product, November~23 2021.
\newblock US Patent 11,179,883.

\bibitem{fischer2013three-}
Joachim Fischer, Jonathan~B Mueller, Johannes Kaschke, Thomas~JA Wolf,
  Andreas-Neil Unterreiner, and Martin Wegener.
\newblock Three-dimensional multi-photon direct laser writing with variable
  repetition rate.
\newblock {\em Optics express}, 21(22):26244--26260, 2013.

\bibitem{purtov2019nanopillar}
Julia Purtov, Peter Rogin, Andreas Verch, Villads~Egede Johansen, and Ren{\'e}
  Hensel.
\newblock Nanopillar diffraction gratings by two-photon lithography.
\newblock {\em Nanomaterials}, 9(10):1495, 2019.

\bibitem{pingali2022reaction}
Rushil Pingali and Sourabh~K Saha.
\newblock Reaction-diffusion modeling of photopolymerization during femtosecond
  projection two-photon lithography.
\newblock {\em Journal of Manufacturing Science and Engineering},
  144(2):021011, 2022.

\bibitem{guney2016estimation}
MG~Guney and GK~Fedder.
\newblock Estimation of line dimensions in 3d direct laser writing lithography.
\newblock {\em Journal of Micromechanics and Microengineering}, 26(10):105011,
  2016.

\bibitem{palmer2021simulation}
Thomas Palmer, Erik~H Waller, Heiko Andra, Konrad Steiner, and Georg von
  Freymann.
\newblock Simulation model for direct laser writing of metallic microstructures
  composed of silver nanoparticles.
\newblock {\em ACS Applied Nano Materials}, 4(9):8872--8879, 2021.

\bibitem{adao2022two}
Ricardo~MR Ad{\~a}o, Tiago~L Alves, Christian Maibohm, Bruno Romeira, and
  Jana~B Nieder.
\newblock Two-photon polymerization simulation and fabrication of 3d
  microprinted suspended waveguides for on-chip optical interconnects.
\newblock {\em Optics Express}, 30(6):9623--9642, 2022.

\bibitem{liu20183d}
Xiaojiang Liu, Hongcheng Gu, Min Wang, Xin Du, Bingbing Gao, Abdelrahman Elbaz,
  Liangdong Sun, Julong Liao, Pengfeng Xiao, and Zhongze Gu.
\newblock 3d printing of bioinspired liquid superrepellent structures.
\newblock {\em Advanced Materials}, 30(22):1800103, 2018.

\bibitem{2020SciPy}
Pauli Virtanen, Ralf Gommers, Travis~E. Oliphant, Matt Haberland, Tyler Reddy,
  David Cournapeau, Evgeni Burovski, Pearu Peterson, Warren Weckesser, Jonathan
  Bright, St{\'e}fan~J. {van der Walt}, Matthew Brett, Joshua Wilson, K.~Jarrod
  Millman, Nikolay Mayorov, Andrew R.~J. Nelson, Eric Jones, Robert Kern, Eric
  Larson, C~J Carey, {\.I}lhan Polat, Yu~Feng, Eric~W. Moore, Jake
  {VanderPlas}, Denis Laxalde, Josef Perktold, Robert Cimrman, Ian Henriksen,
  E.~A. Quintero, Charles~R. Harris, Anne~M. Archibald, Ant{\^o}nio~H. Ribeiro,
  Fabian Pedregosa, Paul {van Mulbregt}, and {SciPy 1.0 Contributors}.
\newblock {{SciPy} 1.0: Fundamental Algorithms for Scientific Computing in
  Python}.
\newblock {\em Nature Methods}, 17:261--272, 2020.

\bibitem{hering2016automated}
Julian Hering, Erik~H Waller, and Georg Von~Freymann.
\newblock Automated aberration correction of arbitrary laser modes in high
  numerical aperture systems.
\newblock {\em Optics Express}, 24(25):28500--28508, 2016.

\bibitem{nelder1965simplex}
John~A Nelder and Roger Mead.
\newblock A simplex method for function minimization.
\newblock {\em The computer journal}, 7(4):308--313, 1965.

\bibitem{Matlab}
{The MathWorks, Inc.}
\newblock corr2.
\newblock {https://www.mathworks.com/help/images/ref/corr2.html}.

\bibitem{ISO25178-603}
{International Organization for Standardization ISO}.
\newblock {Geometrical product specifications (GPS) — Surface texture: Areal
  — Part 603: Nominal characteristics of non-contact (phase-shifting
  interferometric microscopy) instruments}.
\newblock Standard, {International Organization for Standardization ISO},
  Geneva, CH, 2013.

\bibitem{ISO25178-70}
{International Organization for Standardization ISO}.
\newblock {Geometrical product specification (GPS) — Surface texture: Areal
  — Part 70: Material measures}.
\newblock Standard, {International Organization for Standardization ISO},
  Geneva, CH, 2014.

\bibitem{NanoGuide}
{Nanoscribe GmbH}.
\newblock {\em NanoGuide -- Applications Overview}.
\newblock Nanoscribe GmbH.

\end{thebibliography}

\bibliographystyle{unsrt}

\end{document}